\documentclass[review]{elsarticle}
\usepackage{hyperref}
\usepackage{graphicx}
\usepackage{subfig}

\journal{Journal of \LaTeX\ Templates}

\usepackage{numcompress}

\begin{document}

\begin{frontmatter}

\title{Operation and readout of the CGEM Inner Tracker }


\author[mymainaddress,mysecondaryaddress]{Ilaria Balossino\fnref{myfootnote}} 
\fntext[myfootnote]{on behalf of the working group}


\address[mymainaddress]{Istituto Nazionale di Fisica Nucleare, Sezione di Ferrara, Italy}
\address[mysecondaryaddress]{Institute of High Energy Physics, Chinese Academy of Sciences, PRC}

\begin{abstract}
A recently approved ten-year extension of the BESIII experiment (IHEP, Beijing) motivated an upgrade program for both the accelerator and the detector. In particular, the current inner drift chamber is suffering from aging and the proposal is to replace it with a detector based on the cylindrical GEM technology.
The CGEM inner tracker (CGEM-IT) consists of three coaxial layers of triple GEM. The tracker is expected to restore efficiency, improve z-determination and secondary vertex position reconstruction with a spatial resolution of 130$\,$$\mu$ m in the xy-plane and better than 300$\,$$\mu$ m along the beam direction.
A dedicated readout system was developed. Signals from the detector strips are processed by TIGER, a custom 64-channel ASIC that provides an analog charge readout via a fully digital output up to about 50$\,$fC, less than 3$\,$ns jitter. TIGER continuously streams over-threshold data in trigger-less mode to an FPGA-based readout module, called GEM Read Out Card, that organizes the incoming data by building the event packets when the trigger arrives.
Two of the three layers are in operation in Beijing since January 2020 remotely controlled. Due to the pandemic situation the integration activity has been continued on a small-scale prototype. Recently, a test beam has been performed at CERN with the final electronics configuration.
In this presentation, the general status of the CGEM-IT project will be presented with a particular focus on the results from the test beam data acquisition.
\end{abstract}


\end{frontmatter}


\section{Introduction}

\paragraph{\textbf{Where} the experiment is} The Institute of High Energy Physics, located in Beijing (P.R.C.), is the biggest laboratory for the study of particle physics at the Chinese Academy of Sciences. The facility includes an electron-positron collider (Being Electron-Positron Collider II, BEPCII). It runs with an energy in the center of mass in the range [2.00, 4.95] GeV and a designed luminosity of 10$^{33}$\,cm$^{-2}$\,s$^{-1}$. At its interaction point the spectrometer (BEijing Spectrometer III, BESIII \cite{ABLIKIM2010345}) can be found. It has a cylindrical symmetry around the beam pipe and it is composed by different detection systems for the identification of the trajectories of the charged particles and the reconstruction of their momentum (Multilayer Drift Chamber), for the measurements of the energy loss of both charged and neutral particles (ElectroMagnetic Calorimeter) and for performing particle identification (Time Of Flight and MUon Counter), all immersed in a 1 T magnetic field. 
The goal of the experiment is the investigation of the physics of the charmonium spectrum to deepen the studies of hadron physics \cite{Ablikim2020}. This can be done with unprecedented precision thanks to the unique features of this facility that allows measuring the cross sections of very different final states, extracting the different line-shapes and collecting large samples of data. 

\paragraph{\textbf{What} stimulated this project} Both accelerator and experiment are undergoing an upgrade organized in different steps to continue the physics program since it will have additional ten years of data taking. For the accelerator different actions have been taken such as increasing the energy of the center of mass, moving to top-up injection to increase the integrated luminosity, and in the next future, the radio-frequency cavities will be improved. In the past years, some sub-detectors of the spectrometer were improved as well. The renovation of interest in this document revolves around the innermost part of the multilayer drift chamber. It is suffering from aging with a gain-loss per year of about 4$\%$ and the collaboration decided to replace it. The Italian group proposed to substitute it with a \textit{Cylindrical Gas Electron Multiplier} composed of three independent triple-GEM detectors. 

\begin{figure}[h!]
    \centering
    \includegraphics[width=0.45\textwidth]{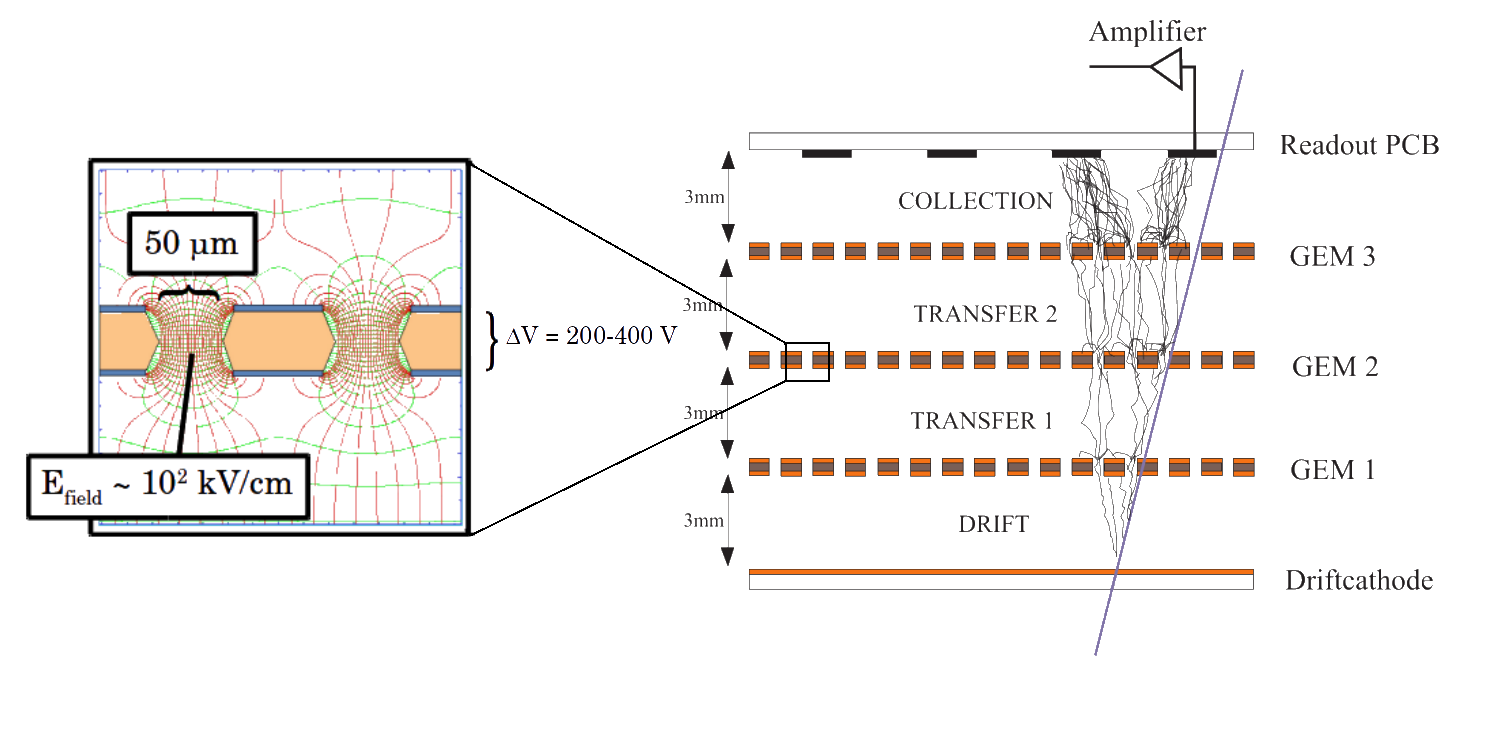} 
    \caption{Schematic of a triple-GEM detector with a simulation of the movement of the electrons created by the primary ionization then multiplied in three stages. The intense electric field inside the holes is made visible in the zoom.}
    \label{fig:gem}
\end{figure}

\paragraph{\textbf{Why} GEM technology was chosen} Since its invention in 1997 \cite{SAULI1997531}, the Gas Electron Multipliers technology has been widely used in several applications. In high energy physics, it allows detecting the interacting charged particles by means of the primary ionization. In detail, a GEM is a Kapton\footnote{Kapton: polyimide film.} foil 50 $\mu$ m thick with 5 $\mu$ m copper faces and pierced with a high density of bi-conical holes (50$\,\mu$ m$/70\,\mu$ m inner/outer diameter and the pitch of about 140$\,\mu$ m). One or more GEM foils are placed between cathode and anode and immersed in a gas mixture (for this application Ar:iC$_4$H$_{10}$, 90:10, or Ar:CO$_2$, 70:30). The electrons created by the ionization are driven towards the anode and pass through the GEM holes. Thanks to the high voltage applied between the copper layers, when the electrons pass through the holes, multiplication occurs due to secondary ionization (Fig.~\ref{fig:gem}). The multi-stage amplification guarantees a high gain (around $10^4$) with low discharge probability compared to other gaseous detectors~\cite{BACHMANN2002294}. These characteristics are the very fruitful ground that makes this detector possible to feature high rate capability, high radiation hardness, and an easily scalable and flexible geometry. \\

\begin{figure}[h!]
    \centering
	\subfloat[][\centering Internal geometry]{\includegraphics[width=0.22\textwidth]{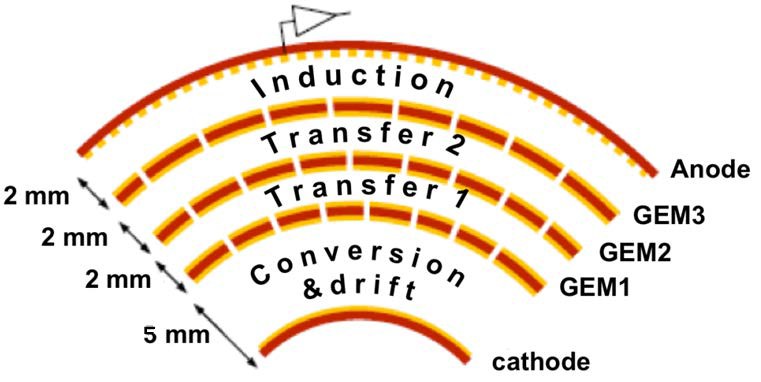}}
    \qquad
    \subfloat[][\centering Readout segmentation]{\includegraphics[width=0.21\textwidth]{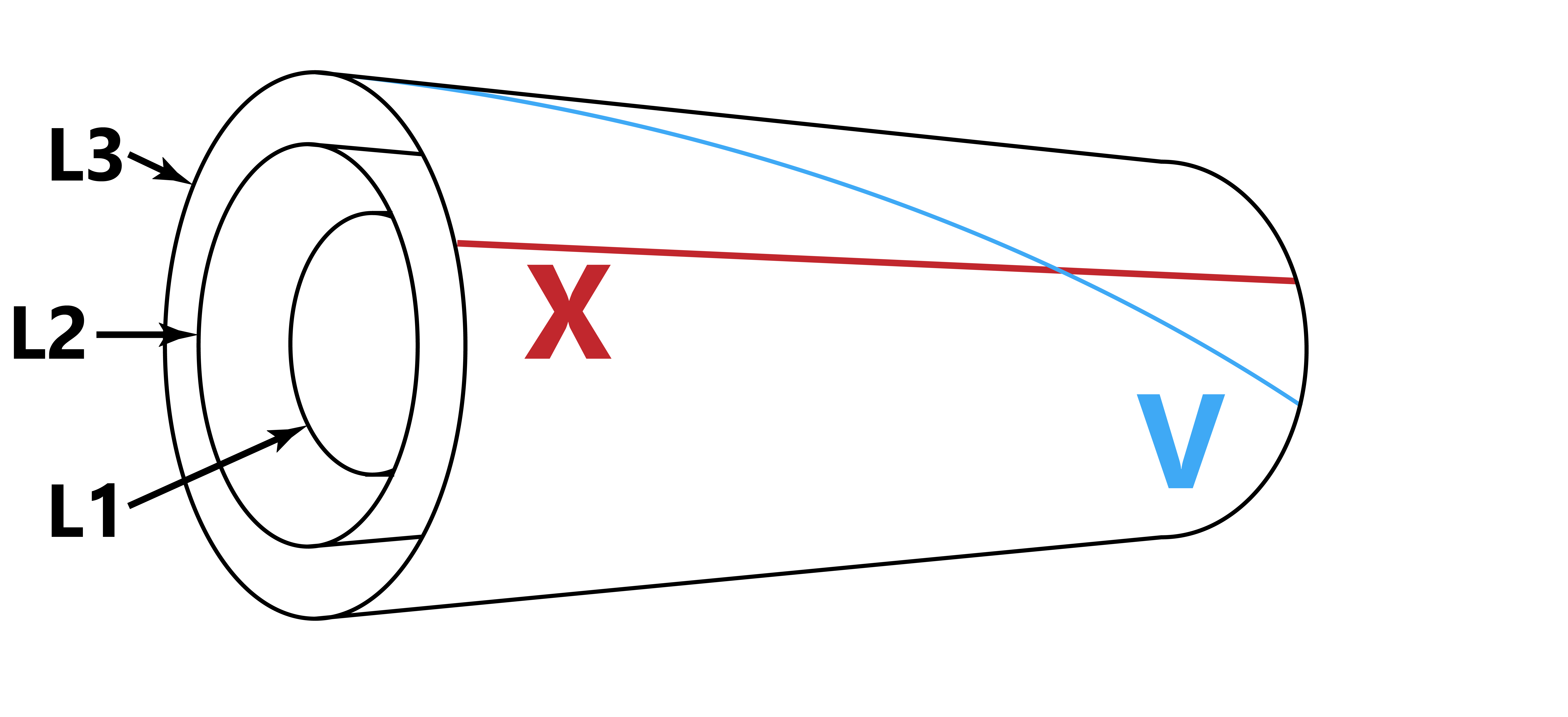}}
    \caption{Schematic of the main features of the cylindrical triple-GEM.}
    \label{fig:cgemit}
\end{figure}

\paragraph{\textbf{Who} is the protagonist} The CGEM Inner Tracker (CGEM-IT) is designed with three layers of cylindrical triple-GEM. Each cylinder, as shown in Fig.~\ref{fig:cgemit}, foresees a cathode in the inner part, three GEM foils, and an anode hosting the readout plane segmented with longitudinal (X) and stereo strips (V). The chosen gaps are respected thanks to the Permaglass\footnote{Permaglass: fiberglass reinforced epoxy resin} rings at the extremities of the each detector. The mechanical support of the cylinder is given by two cylindrical structures: one inside the cathode and the other outside the anode originally studied to be made of a ROHACELL\footnote{ROHACELL\textregistered: light polymethacrylimide based structural foam}-Kapton sandwich and then upgraded to a sandwich of laminated carbon fiber-Honeycomb\footnote{Honeycomb: light core material based on an hexagonal cell geometry}. This design keeps the material budget within the 1.5$\%$ X$_0$ limit while covering a the 93$\%$ of the solid angle. The readout chain has been developed specifically for this project and gives information on the time and the collected charge. The expected performance of this detector ($\sigma_{r\phi} \leq 130$ $\mu$ m; $\sigma_{z} \leq 1$ mm; dp/p @ 1 GeV/\textit{c} $\sim$0.5$\%$) aims to improve the position resolution and the reconstruction of the secondary vertices.

\section{Development}
The project is still ongoing and the development of the main activities with which the working group is engaged will be presented. 
\paragraph{Readout Chain} The description of the full chain can be divided into on- and off-detector \cite{Amoroso2021}. Right on the anode ends of each detector, the front-end boards (FEB) are placed. Each board hosts two chips TIGER (Torino Integrated Gem Electronics for Readout) for a total of 128 channels. They can be operated in Sample-and-Hold or Time-Over-Threshold mode to provide charge and time information. The fully-digitized data are then transmitted to the GEMROCs (GEM Read-Out Cards). Here, the instruction to power, monitor, and configure are given and the trigger-matching operations are performed. The timing of the system is guaranteed by the Fast Control System Fanout and Local Fanout. Once in the experiment, the readout chain will be completed by two Data Concentrator boards to build the events and store them in the BESIII Data Acquisition System. 
Commissioning involves different tests, debugs, and characterizations from the single-channel noise test to the firmware optimization. A Graphical User Front-end Interface (GUFI) has been coded in Python with tools to configure, test, and run acquisition in a stand-alone setup. It also includes a monitoring for the main parameters of the readout chain such as the temperature and the voltages. 

\begin{figure}[h!]
    \centering
    \subfloat[][\centering Charge centroid method]{{\includegraphics[width=0.2\textwidth]{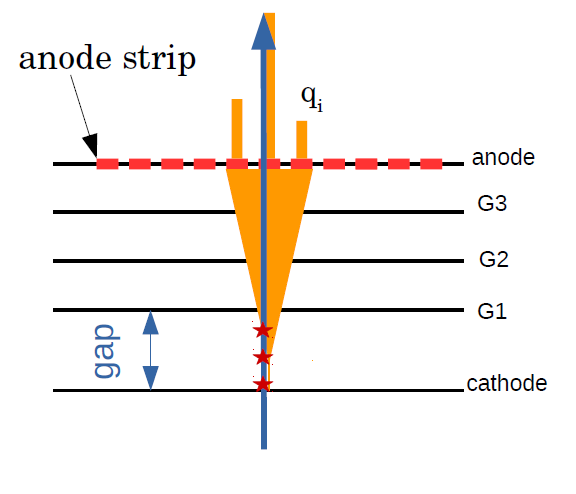} }\label{fig:recCC}}
    \qquad
    \subfloat[][\centering $\mu$ TPC method]{{\includegraphics[width=0.22\textwidth]{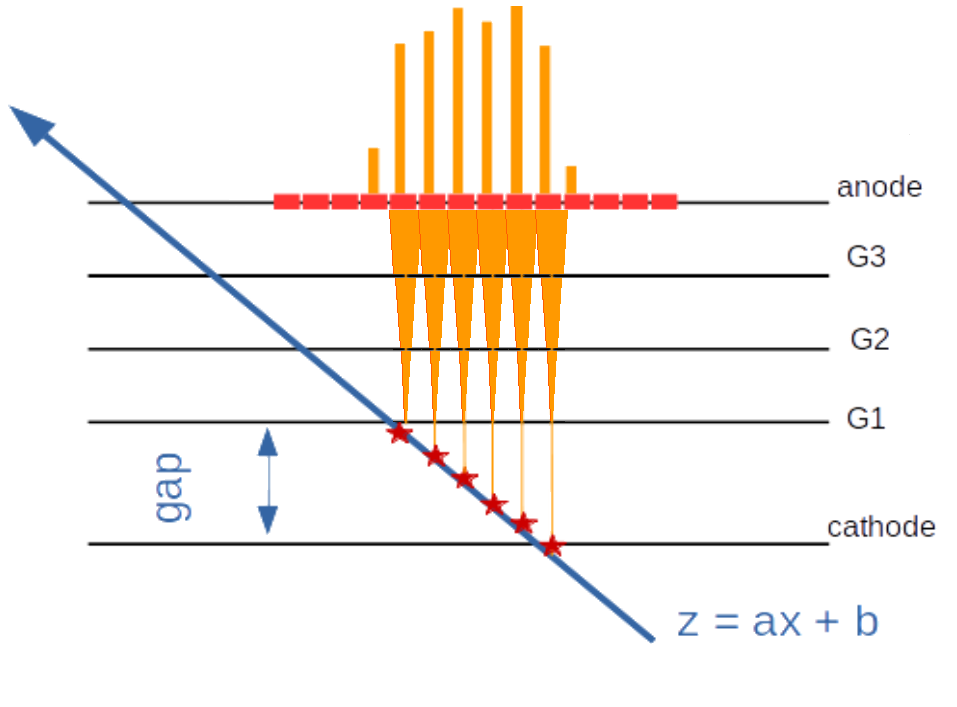} }\label{fig:recuTPC}}
    \caption{Sketch of the reconstruction algorithms. At each red star corresponds one ionization point; the electron avalanche is then represented with the orange triangles crossing the gaps and the sub-layers; the segmented anode is represented with the red dashes and the distribution of the induced charge at the anode is represented on top of them.}%
\end{figure}

\paragraph{Software} This work can be divided into two parts as well: the first is the preparation and optimization of the CGEM-IT software to be included in the BESIII Offline Software System; the second is specifically focused on the position reconstruction algorithms. 
The offline software system provides the description of the geometry of the CGEM-IT together with the simulation and the reconstruction of the real data to measure the particle path inside this detector. This is performed with global track finder algorithms and track fitting procedure. Up to now, calibration and alignment are ongoing and then this code will be included in the final framework.
The position reconstruction, one of the flagship of this project, is performed with two methods merged together considering that the charge of all the strips of the cluster is proportional to the deposited energy and to the geometrical position of the ionization inside the drift gap: Charge Centroid (CC, Eq.~\ref{eq:recCC}, Fig.~\ref{fig:recCC}) and micro$\,$-$\,$Time Projection Chamber ($\mu$ TPC, Eq.~\ref{eq:recuTPC}, Fig.~\ref{fig:recuTPC}). 

\vspace{5mm}
\begin{minipage}[b]{0.45\linewidth}
\begin{equation}
 x_{CC} = \frac{\Sigma_i^{N_{hit}} Q_{hit,i}\, x_{hit,i}}{\Sigma_i^{N_{hit}}Q_{hit,i}}
\label{eq:recCC}
\end{equation}
\end{minipage}
\begin{minipage}[b]{0.45\linewidth}
\begin{equation}
  x_{\mu TPC} = \frac{gap/2 - b}{a}
\label{eq:recuTPC}
\end{equation}
\end{minipage}
\vspace{5mm}

The CC averages the charge of all the strips of the cluster by weighting it by its charge. The $\mu$ TPC instead considers the drift gap as a tiny TPC and with position and time information, it associates each strip with a bi-dimensional point and uses a linear fit to extrapolate the particle position. This merging is useful to optimize the reconstruction at different angles in a magnetic field of 1 T. When the angle between the incident particle and the CGEM surface is small, the charge distribution is Gaussian and the CC method is more efficient. As the angle increases, this distribution is no longer Gaussian and the $\mu$ TPC method becomes more adaptive to optimize the resolution.

\begin{figure}[h!]
    \centering
    \subfloat[][\centering Final layers]{{\includegraphics[width=0.19\textwidth]{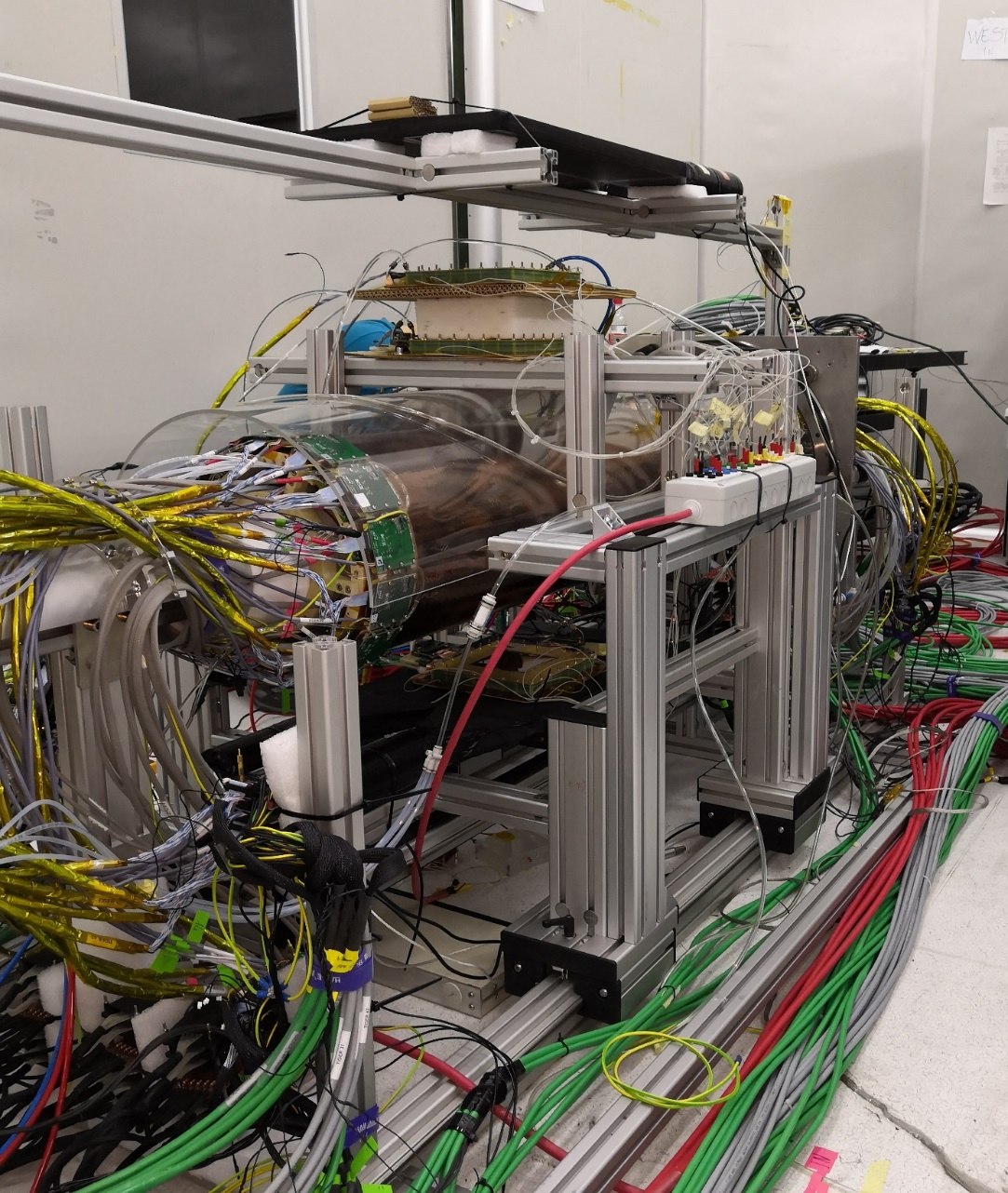}}\label{fig:cos}}
    \qquad
    \subfloat[][\centering Prototypes]{{\includegraphics[width=0.202\textwidth]{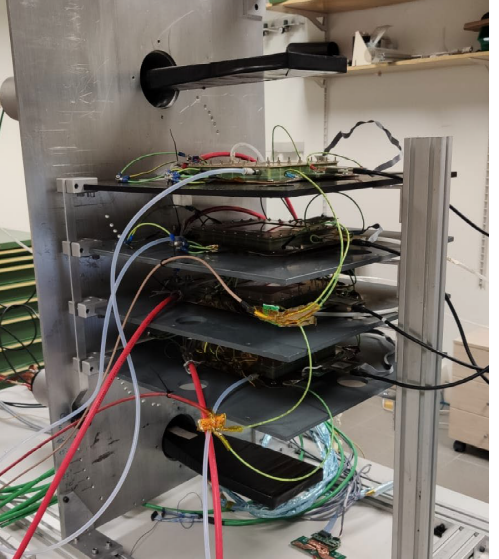}}\label{fig:beam}}
    \caption{Data taking setup.}
    \label{fig:data}
\end{figure}

\paragraph{Data Taking} The commissioning will be completed only when the full CGEM-IT will be assembled together and will have collected a satisfying amount of cosmic data to prove the expected performance. Up to now, the commissioning is moving on two different paths forced by the impossibility to travel in China in the past years: the first, Fig.~\ref{fig:cos}, in Beijing, with two (out of three) final layers instrumented with the full readout chain; the second, Fig.~\ref{fig:beam}, in Italy, with planar detectors to continue the electronics/detector integration and optimization. 
At IHEP, the two cylindrical GEMs are assembled on top of the assembly machine and equipped with cables, readout cards, and a power supply system, mimicking the final conditions in the experiment. The data taking started at the end of 2019 and it has been used for the development of the reconstruction software. Since January 2020, the system is controlled remotely with extraordinary on site maintenance and a remote shift system has been organized to monitor the detectors while acquiring data to guarantee the safety of the detector since a complete interlock system was not organized yet. 
In Italy, in 2021, a stand-alone setup with four planar triple-GEM prototypes has been implemented to be used as a telescope for cosmic data while in Ferrara and for a two weeks test beam data taking in H4 line at CERN. From this setup, the first sought results focused on a comparison of two readout chains: APV+SRS, often used for MPGD detectors and previously used to validate triple-GEMs in magnetic field~\cite{Alexeev:2019rng}; TIGER+GEMROC, specifically developed for the CGEM-IT. Now, the setup has been rebuilt in Ferrara and is fully dedicated to the CGEM-IT readout chain, the electronics/detector integration continues with online and offline analysis and, when necessary, hands-on tests.

\paragraph{Construction} The third, outer-most, detector that composes the CGEM-IT has not been constructed yet. Thorough investigation and testing are underway to ensure the proper rigidity to the detector without exceeding in the material budget, and considering the necessary handling and movement before the final installation in the experiment. The design validation is ongoing by means of electrical tests, x-rays and CT scans, and FEM analysis. The goal is to provide answers as precise as possible to define the optimal design for this detector. Preliminary results have been presented at \cite{Balossino_2020}. 

\section{Results}

\begin{figure}[h!]
    \centering
    \includegraphics[width=0.4\textwidth]{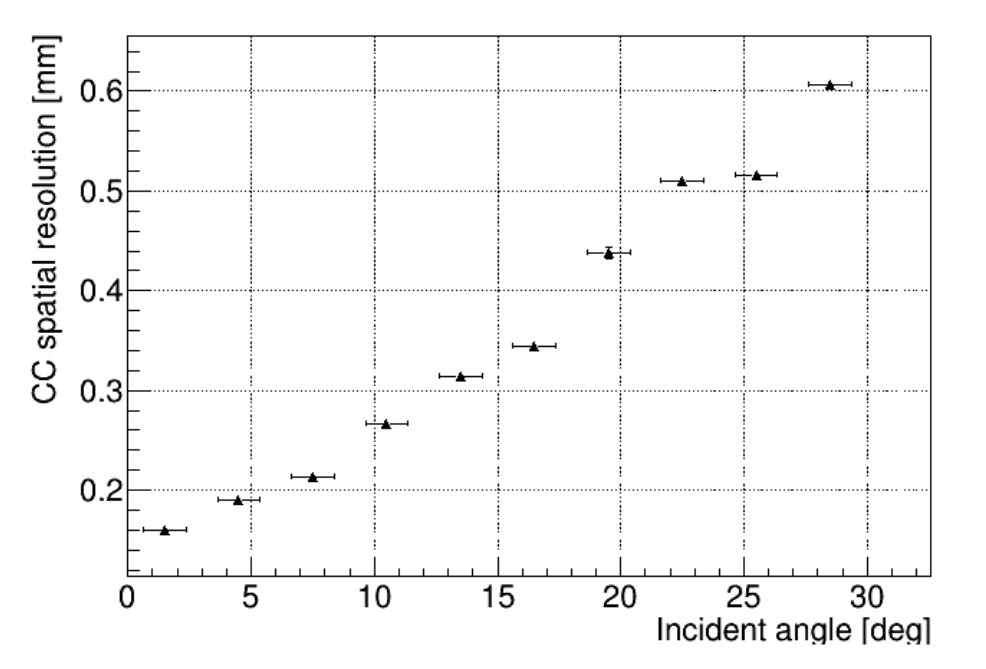} 
    \caption{Charge centroid spatial resolution as a function of the angle between the incident charged particle and the surface of the detector.}
    \label{fig:cc}
\end{figure}

With two out of three final detectors in a cosmic stand in Beijing operating remotely some preliminary reconstruction analyses have been performed~\cite{Farinelli_2020}. Fig.~\ref{fig:cc} shows the spatial resolution extracted from the charge centroid algorithm with respect to the angle between the incident particle and the detector surface. It can be seen that the measurements follow the expected performance by degrading as the incident angle increases.

\begin{figure}[h!]
    \centering
    \includegraphics[width=0.4\textwidth]{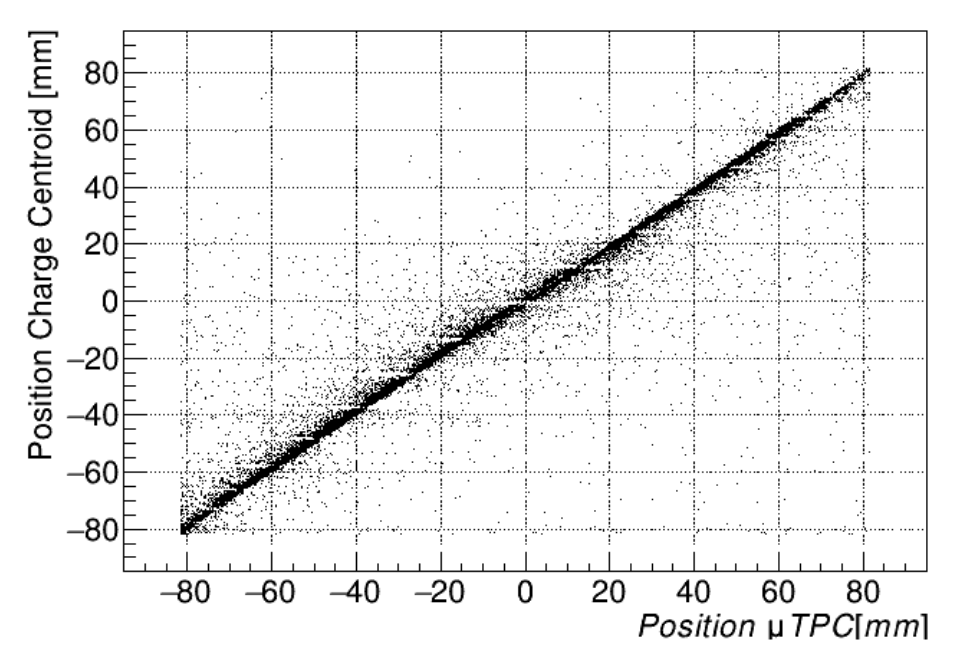} 
    \caption{Comparison of the reconstructed position extracted with the charge centroid algorithm and the $\mu$ TPC algorithms.}
    \label{fig:tpc}
\end{figure}

A second preliminary result extracted from the setup in Beijing presented in Fig.~\ref{fig:tpc} is the first implementation of the $\mu$ TPC algorithm prepared for the CGEMBOSS framework compared with the charge centroid one. It shows a good correlation between the two algorithms and it is promising for future implementations of the code. 
The data taking is still ongoing and preliminary analyses point to good stability in time.

\begin{figure}[h!]
    \centering
    \subfloat[][\centering Resolution wrt GEMS's HV]{{\includegraphics[width=0.22\textwidth]{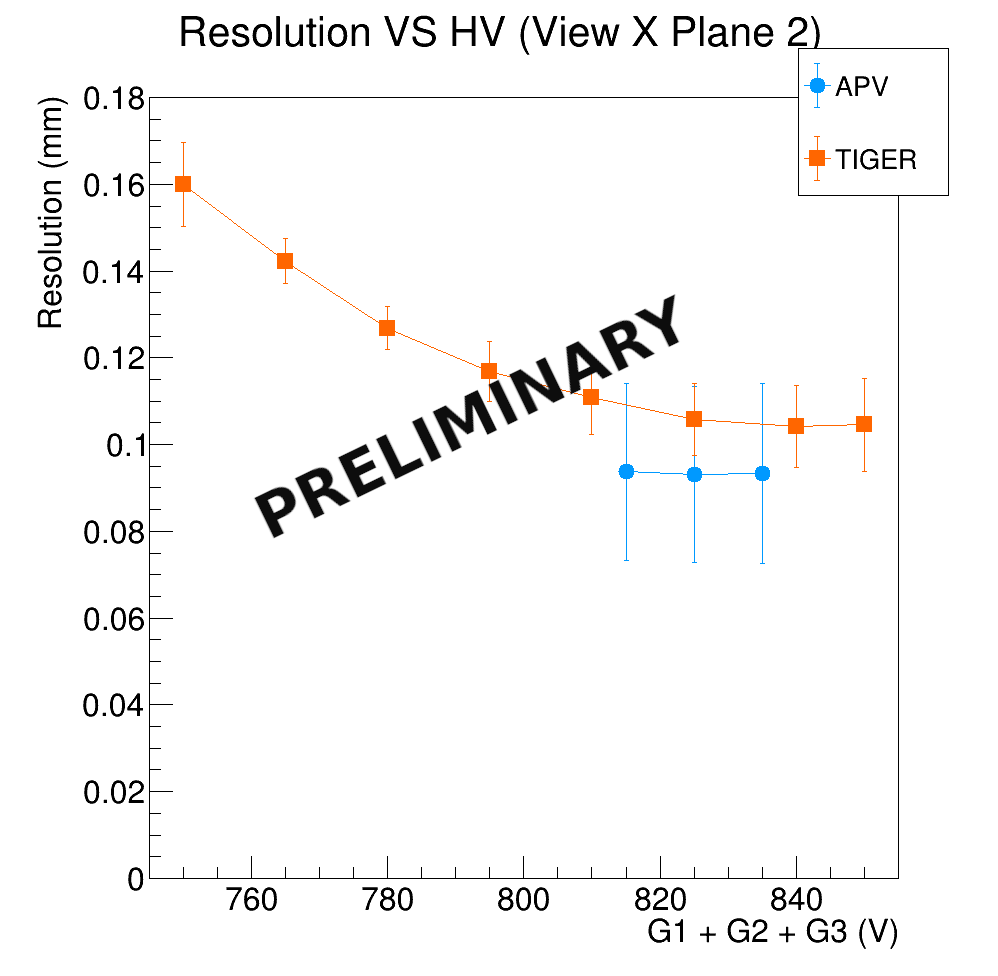}}\label{fig:res1}}
    \qquad
    \subfloat[][\centering Cluster size wrt GEM's HV]{{\includegraphics[width=0.22\textwidth]{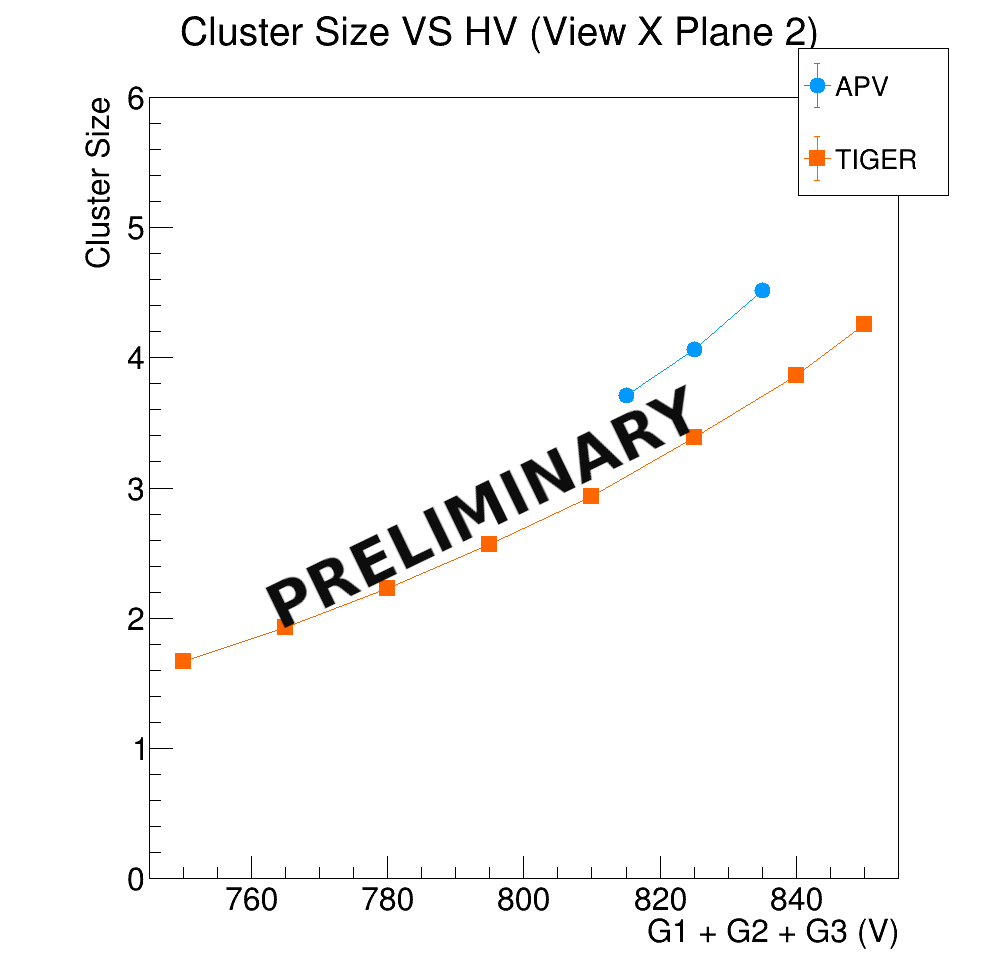}}\label{fig:res2}}
    \caption{Preliminary results of the comparison between the APV+SRS and TIGER+GEMROC readout chains. }
    \label{fig:restb}
\end{figure}

From the test beam and cosmic stand-alone setup with the planar GEMs the work started with a comparison of two readout chains: APV+SRS and TIGER+GEMROC. Some examples presented in Fig.~\ref{fig:restb} show that, within the errors, the results can be considered compatibles. The work then continued with the CGEM-IT electronics with scans of the main parameters such as high voltage, incident angle, threshold, or integration time and the analysis is still ongoing.

\section{Conclusions}
The CGEM-IT project involves the construction and commissioning of an inner tracker for the BESIII experiment built of three co-axial cylindrical triple-GEM detectors. Each cylinder is an independent detector that can reconstruct the 3D position of an incident particle thanks to its segmented anode. 
Two of three detectors have already been built and shipped to Beijing.
The third has yet to be built as an investigation is underway to optimise the design due to its fragility and its necessary handling until it is installed. 
A data taking is ongoing remotely since January 2020 with two cylinders assembled together in a temporary cosmic stand with the final readout chain. It is showing promising results and stability in time, and the optimization of the setup will be performed as soon as the full Italian team of researchers and experts is able to travel again. 
In the meantime, in Italy, a stand-alone setup has been implemented with four planar triple-GEM prototypes instrumented with the final electronics of the CGEM-IT. This has been done to continue the optimization of the electronics/detector integration that, once we will be able to travel again, will be implemented in the final system to finalize the commissioning.

<inserted text>
 \section*{Acknowledgments}
This research was funded by the European Commission in the RISE Project 645664-1244
BESIIICGEM, H2020-MSCA-RISE-2014 and in the RISE Project 872901-FEST, H2020-MSCA-RISE-1245
2019

\bibliography{mybibfile}

\end{document}